\documentclass[12pt]{iopart}
\usepackage[english]{babel}
\usepackage{bm,amssymb,amsfonts}
\usepackage{enumerate}
\usepackage{graphicx}

\usepackage[latin3]{inputenc}

\newtheorem{remark}{Remark}
\newtheorem{theorem}{Theorem}

\newtheorem{proposition}{Proposition}
\newtheorem{definition}{Definition}

\newcommand{\al}{\alpha}
\newcommand{\be}{\beta}
\newcommand{\la}{\lambda}

\newcommand{\beq}{\begin{equation}}
\newcommand{\eeq}{\end{equation}}
\newcommand{\bea}{\begin{eqnarray}}
\newcommand{\eea}{\end{eqnarray}}
\newcommand{\bdef}{\begin{definition}}
\newcommand{\fdef}{\end{definition}}
\newcommand{\bp}{\begin{proposition}}
\newcommand{\ep}{\end{proposition}}
\newcommand{\bt}{\begin{theorem}}
\newcommand{\et}{\end{theorem}}
\newcommand{\ben}{\begin{enumerate}}
\newcommand{\een}{\end{enumerate}}

\newcommand{\ba}{\begin{array}}
\newcommand{\ea}{\end{array}}

\begin{document}
\jl{6}
\title{Reference frames and rigid motions in relativity: Applications}
\author{D. Soler}
\address{ Oinarrizko Zientziak Saila, Goi Eskola Politeknikoa,			
Mondragon Unibertsitatea}
\ead{dsoler@eps.mondragon.edu}
\begin{abstract}
The concept of rigid reference frame and of constricted spatial metric, given in the previous work [\emph{Class. Quantum Grav.} {\bf 21}, 3067,(2004)] are here applied to some specific space-times: In particular, the rigid rotating disc with constant angular velocity in Minkowski space-time is analyzed, a new approach to the Ehrenfest paradox is given as well as a new explanation of the Sagnac effect. Finally the anisotropy of the speed of light and its measurable consequences in a reference frame co-moving with the Earth are discussed.
\end{abstract}

%

\maketitle

\section{Introduction \label{S1}}

In the design of individual experiments,  experimentalists assume, often tacitly, that there exist some objects that are invariant under arbitrary rotations and translations.

Consider for example the experiments where a resonant cavity or  an interferometer is used to detect the anisotropy of the speed of light. Experimentalists use these objects,  made with a material as rigid as possible, as standards of length.

However, there are no real rigid bodies (neither in Relativity theories, nor in  Newtonian mechanics), the real bodies are approximately rigid, and their points approximately follow a {\bf rigid motion}, that is, the motion of an ideal rigid body:
a) whose geometry does not change either during the experiment  or b) when it is moved around.

Briefly, it is assumed that some objects have the intuitive properties of reference spaces in Newtonian mechanics: they are approximately rigid and they  fulfil the axiom of free mobility \cite{Cartan}.

The notion of rigid motion is often related to the existence of a class of congruences admitting a metric ($\bar g$) projectable onto teh reference space(definition \ref{d4}), but the free mobile condition requires (theorem \ref{t2}) $\bar g$ to be maximally symmetric, i.e. to have constant curvature.

In the previous work \cite{LlosaSoler04} we prove  the existence of a wide enough class of congruences admitting a  maximally symmetric projectable metrics ($\bar g$) which can be obtained by a constriction transformation from the quotient metric ($\hat g= g+u\times u$). (Theorem \ref{t3}). 

Any of these congruences can be used to model the kind of motion followed by  real bodies rigid ``enough'', and the corresponding  spatial metric $\bar g$ will  be interpreted as the metric  that is embodied by standard rods in the lab frame.

This election provides an instrument for measuring distances directly, that is, independent of speed of light and, therefore, of the  Einstein convention of simultaneity.  In other words, 
we do not interpret the quotient metric as the space metric.

Even though we have eliminated the interpretation of the quotient metric  as the ``physical spatial metric'', this object has not lost its importance. On the contrary, it can be interpreted as an ``optical metric'', for this reason we shall call this tensor  as Fermat metric. 

The present work is organized as follows: the next section, after recalling some basic definitions of reference frames,  shows our main results obtained in \cite{LlosaSoler04}. The physical interpretation of these results allows us, in section \ref{sec:2}, to study the uniformly rotating rigid disc from a new point of view. We explain new approach to the Ehrenfest paradox, and we compare it with some other explanations available in recently published papers.


The possibility of measuring distances and times with different metrics, $\bar g$ and $\hat g$ respectively, make sense to question the isotropy of the speed of light. In section \ref{sec:3}, this analysis allows us to interpret physically the 1-form $\mu$ appearing in the constriction transformation (remark \ref{r2}).

Finally, we study the anisotropy of the speed of light in a uniformly rotating reference frame, and in  a reference frame co-moving with the Earth, giving new interpretation of the well known Sagnac effect \cite{Sagnac} for the rotating disc case, and a theoretical explanation of the Brillet and Hall experimental results \cite{Brillet}. This is done in a very similar way to Bel et al. \cite{Bel99,  BelMol00, Aguirregabiria, BelMarMol94}, but with different experimental predictions.

\section{Basic Results\label{sec:1}}
\subsection{Space of reference and time scale}

A reference frame consists of a space or reference, a time scale and a spatial metric. Let us define these concepts:

Let ${\cal V}_4$ be a spacetime domain, $g$ a Lorentzian metric with signature $(-+++)$ and $u$ a unitary timelike vector field on ${\cal V}_4$, $\,g(u,u) = -1\,$.
Its integral curves yield a three-parameter congruence of timelike worldlines (a timelike 3-congruence), ${\cal E}_3$, that establishes the following equivalence relation: for $\,x, y \in {\cal V}_4\,,$
$$ x\sim y \Leftrightarrow y \mbox{ lies in the worldline passing through } x \,.$$

The cosets for this equivalence relation are the worldlines belonging to the 3-congruence

\begin{definition}\label{espai}
 The quotient space $\; {\cal E}_3 := {\cal V}_4/\sim \,$
is \textbf{the space of reference} ---briefly, {\em the space}--- of the congruence.
\end{definition}
The canonical projection $\pi:\,{\cal V}_4 \longrightarrow {\cal E}_3$ is a differentiable map, and the tangent and pull-back maps will be denoted, respectively, by  
$$\pi_\ast:\,T{\cal V}_4 \longrightarrow T{\cal E}_3 \qquad  {\rm and}  \qquad \pi^\ast:\,\Lambda{\cal V}_4 \longrightarrow \Lambda{\cal E}_3\,.$$

\begin{definition}
A \textbf{time scale} for the congruence ${\cal E}_3$ in the domain ${\cal V}_4$ is a 1-form $\theta \in\Lambda^1{\cal V}_4$ such that: $\langle\theta,u\rangle \neq 0$ everywhere in ${\cal V}_4$. 
\end{definition}
A curve in ${\cal V}_4$ whose tangent vector $v$ satisfies $\langle\theta, v\rangle =0$ connects events that are simultaneous according to the time scale $\theta$ ---briefly, $\theta$-simultaneous. 

Furthermore, if $\,\langle\theta,u\rangle = -1\,$, then $\theta$ is called a {\it proper time scale}. As a particular case, the {\it Einstein proper time scale} 
$$  \theta_0 := g(u,\_\,)  $$
is the result of combining the Lorentzian metric $g$ and the unitary vector field $u$ that defines the congruence. This time scale is physically based on a system of identical local clocks, ticking local proper time and synchronized according to Einstein's protocol.

\subsection{Rigid reference frame}

A {\it rigid reference frame} is defined by the triple $({\cal E}_3,\theta,g_3)$ ---a timelike 3-congruence, a proper time scale and a  special Riemannian metric on the reference  space. 

A Riemannian metric $g_3$ of positive signature on the reference space ${\cal E}_3$ can be pulled back to a metric $\overline{g}:=\pi^\ast g_3$ on ${\cal V}_4$, so that the $\overline{g}$-metric product between any couple of vector fields tangent to ${\cal V}_4$ is
$$
\overline{g}(v,w) = g_3(\pi_\ast v,\pi_\ast w)\circ \pi
$$

This metric satisfies the following three conditions:
\begin{itemize}
\item The metric $\overline{g}$ is degenerated,
 
\begin{equation}
\overline{g}(u,\_) =0
\label{e1}
\end{equation}
\item the $\overline{g}$-product of any two vector fields $v$ and $w$ on ${\cal V}_4$ is non-negative:
\begin{equation}
\overline{g}(v,w) \geq 0
\label{e2}
\end{equation}
\item it is rigid
\beq \label{e3}
\,{\cal L}(u) \overline{g} =0.
\eeq
\end{itemize}

\begin{definition} \label{d0}
A symmetric 2-covariant tensor field $\overline{g}$ on ${\cal V}_4$ is said to be {\bf a spatial metric} relatively to the congruence ${\cal E}_3$ if it satisfies (\ref{e1}) and(\ref{e2}).
\end{definition}

Taking into account that if a  spatial metric $\overline{g}$ satisfies (\ref{e3}), then it can be unambiguously projected onto a Riemannian metric $g_3$ on ${\cal E}_3$, we can introduce the notion of projectable spatial  metric:

\bdef\label{d4}
A spatial metric  $\overline{g}$ on ${\cal V}_4$ is said to be  {\bf projectable} or {\bf rigid}  relatively to the congruence ${\cal E}_3$ if it satisfies (\ref{e1}),(\ref{e2}) and (\ref{e3}).
\end{definition}

\subsection{Rigid Motion}

A rigid metric $\overline{g}$ can always be associated with a 3-congruence ${\cal E}_3$, at least locally. Indeed, it is sufficient to assign values for $\overline{g}$ on a hypersurface that is nowhere tangential to the worldlines in ${\cal E}_3$ and extend it according to the transport law ${\cal L}(u)\overline{g} =0$. However, a spatial metric like this, would rarely have any physical meaning.

The simplest choice for a physically significant spatial metric is:
\begin{equation}
\overline{g} := \hat{g} \qquad {\rm where} \qquad \hat{g}_{\alpha\beta}:= g_{\alpha\beta} + u_\alpha u_\beta\;,
\label{e5}
\end{equation}
i.e. the quotinet or Fermat metric \cite{Landau}. But, it is a well-known fact that  this metric is rigid only for a very short class of congruences, \cite{Herglotz}. Moreover, this tensor is rarely a constant curvature tensor, so if we take this tensor as a metric tensor for the space of reference, this space will not satisfy the property of free mobility.

As it has been pointed out in the introduction, this desirable property states that rods used as standards for measuring space geometry must be invariant when move around.

This statement is very restrictive when Riemannian manifolds are taken into account because of \cite{Cartan}
\bt\label{t2}[Cartan]
Any Riemannian manifold that admits movements that do not deform the geometry of a figure (satisfies the free mobility axiom), is maximally symmetric, that is, has  constant curvature.
\et

So we think, as well as other authors,\footnote{See for instance \cite{Bel96} or \cite{Bel05} where a  very interesting discussion on physical consequences of imposing or not to  $\bar g$ to have constant curvature  can be found.} that apart from  conditions (\ref{e1}-\ref{e3}) it is necessary that the projectable metric $\overline g$  has constant curvature.

In \cite{LlosaSoler04}, a frame connection ($\overline \nabla$), that  is a spacetime connection adapted to the congruence and to a spacial metric, is given, and it is shown that  when a spatial metric is projectable, the projection of the frame connection is the Riemannian connection associated to the projected metric.

Therefore, the constant curvature conditions yields to the equation 
      \begin{equation}\label{e4}
       \overline{K}_{\mu\nu\alpha\beta } = k \,(\overline{g}_{\mu\al}\,\overline{g}_{\nu\be} - \overline{g}_{\mu\be}\,\overline{g}_{\nu\al})
       \end{equation}
      where $k$ is a constant and $\overline{K}_{\mu\nu\alpha\beta }$ is the covariant curvature tensor\footnote{This curvature tensor is related to other curvature tensors for the space of reference \cite{Catt, Zelm, Boers95, RizziRuggiero02}. In \cite{tesi} there is a extensive discussion on this subject and  relations between them are given.} of the frame connection $\overline \nabla$.

After these remarks we give the following      
      
\bdef\label{d5}
A \textbf{Rigid Motion}  is the triple $({\cal E}_3,\theta,\bar g)$ such that (\ref{e1}), (\ref{e2}), (\ref{e3}) and (\ref{e4}) are fulfilled. 
\fdef

Notice that even when the Fermat tensor is a projectable metric, that is, even when the 3-congruence is a Born congruence \cite{Born}, this tensor do not represent, in general, the standards rods used in a laboratory,  whereas  it is necessary to deform it and obtain a new spatial metric, projectable and \textbf{maximally symmetric}. In particular, our aim is to find a slight variation of formula (\ref{e5}) that relates $\bar g$ with $\hat g$.

We propose that relation to be a  {\bf constriction}, that is, at every point, there is a conformal transformation  on the orthogonal plane of an space direction given by $\mu$, this yields to the relation
\begin{equation}
\overline{g} = \varphi\,(\hat g + \epsilon\,\mu\otimes\mu) \;,\qquad\qquad \epsilon=\pm 1
\label{e3.2}
\end{equation}
where $\mu$ is a 1-form orthogonal to $u$ and $\varphi$ is a positive function.


\begin{figure}[ht!]
	  \begin{center}
		\includegraphics[scale=0.45]{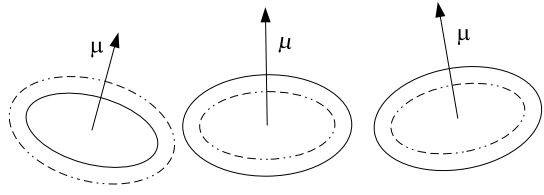}
		\end{center}
		\caption{At every point, there is a conformal transformation  on the orthogonal plane of an space direction given by $\mu$.}
	
\end{figure}

In this context, the following theorem proved in \cite{LlosaSoler04} states that, in any spacetime, it  is always possible to find  rigid motions, and we interpret the constant curvature projectable metric $\bar g$ as the metric that is embodied by standard rods used in a laboratory. 

\bt\label{t3}
Given  a spacetime with constant signature $({\cal V}_4,g)$, it is possible to find wide enough  class of reference frames $({\cal E}_3,\overline{g},\theta:=g(u,\_))$ on ${\cal U}\subset{\cal V}_4$, such that 
\ben[a)]
\item The metric $\overline{g}$ is related to the Fermat metric by the {\it constriction}:
      \begin{equation}
      \overline{g}_{\alpha\beta } = \varphi\,(g_{\alpha\beta} + u_\alpha u_\beta + \epsilon\,\mu_\alpha\mu_\beta )\,
       \label{e8}
      \end{equation}
      where $\varphi$ and $\|\mu\|^2:=g^{\alpha\beta}\mu_\alpha\mu_\beta\,$ fulfil a previously chosen arbitrary constraint: 
      \begin{equation}
      \Psi(\varphi, \|\mu\|^2 )=0 \,.
      \label{e3.8a}
      \end{equation}
      \item The metric $\overline{g}$ is rigid:
      \begin{equation}
       {\cal L}(u)\bar g_{\al\be} = 0
       \label{e3.6}
      \end{equation}
\item and has constant curvature
      \begin{equation}
       {\cal F}_{\mu\nu\alpha\beta } := \overline{K}_{\mu\nu\alpha\beta } + k \,\overline{g}_{\mu\nu\alpha\beta } = 0
       \label{e3.7}
      \end{equation}
      where $k$ is a constant and 
      \begin{equation}
       \overline{g}_{\mu\nu\alpha\beta }  :=\overline{g}_{\mu\al}\,\overline{g}_{\nu\be} - \overline{g}_{\mu\be}\,\overline{g}_{\nu\al}
       \label{e3.8}
      \end{equation}
\een
\et
 
Remarkably, Born congruences, which have traditionally been used to characterize rigid motions, also provide rigid frames according to this new definition.

\bt Any Born congruence defines a Rigid motion.
\et

\section{Rotating disc\label{sec:2}}
Let $ds^2$  be the line element of Minkowski's spacetime in cylindric coordinates

\bea
ds^2 = -dT^2+dR^2+R^2d\Theta^2+dZ^2 
\eea
the motion of a reference frame with constant angular velocity $\omega$ is a timelike killing congruence with parametric equations: 
\bea \label{8.5}
\{T = t,\ R= r,\ \Theta = \theta+ \omega t,\ Z = z\}
\eea

The unitary tangent  vector is
\beq
u^{\alpha} = \frac{1}{\sqrt{1-\omega^2r^2}}(1,0,\omega,0)
\end{equation}

Using the system of adapted coordinates to the rotating frame of reference, the line element of the spacetime becomes

\bea
ds^2 = -(1-r^2\omega^2)dt^2+2r^2\omega d\theta dt+ dr^2 + r^2d\theta^2+dz^2 \label{8.6}
\eea

The Fermat tensor is, in these coordinates, 
\bea\label{e17}
\hat{g} = dr^2 +\frac{r^2}{1-\omega^2r^2}d\theta^2+dz^2 \label{15}
\eea
so, introducing the  1-form 
\beq \label{8.8}
\mu = \frac{\omega r^2}{\sqrt{1-\omega^2r^2}}d\theta
\end{equation}
and taking $\epsilon = -1$, it is obvious that the $ \overline{g}_{\alpha\beta } = \varphi\,(g_{\alpha\beta} + u_\alpha u_\beta+\epsilon\,\mu_\alpha\mu_\beta )
$
 is a metric with constant curvature.

\beq\label{e16}
\bar g := \hat g -\mu\mu =  dr^2 +r^2d\theta^2+dz^2 \label{16}
\eeq

This case corresponds to choosing the arbitrary  function (\ref{e3.8a})
as  $\Psi = g(\mu , \mu)-\omega^2r^2$.

\subsection{Ehrenfest Paradox\label{sec:2.1}}

The formulation of the Ehrenfest paradox  is   the following \cite{Ehrenfest}:

Consider a disc that spins with respect to an inertial frame of reference (K), with constant angular velocity ($\omega$) around its axis of symmetry. Let  K' be an observer at rest in the disc, and  R and R' respectively the radius  measured from K and K'.

Then a contradiction arises: With respect to K, each element
of the circumference moves in its own direction with instantaneous speed $%
\omega R$, thus the length (L) of the circumference must be affected by the Lorentz contraction, being  $L = \gamma^{-1} L'$. However, considering an element of the radius, its instantaneous velocity is perpendicular to its length; thus, an
element of the radius cannot show any contraction with respect to the rest state. Therefore $R=R^{\prime}$. 

If $L'= 2\pi R'$, then  $L\neq 2\pi R$ and conversely. Hence, Euclidean geometry can hold in only one frame, either ${\cal K}$ or ${\cal K}^\prime$.

Since 1909, there has been a vast literature devoted to this paradox, but according to recent published papers,  it seems that the physics involved in a rigid rotating disc is not clear at all.

In any case, it is clear that a rotating platform defines a non-time-orthogonal physical frame; being the usual 3+1 foliation of the spacetime, meaningless. The splitting  of the spacetime into a space and a time relatively to the reference frame must be that introduced in section \ref{sec:1} (definition \ref{espai}). Therefore, by definition, the disc does not deform.

In this sense, we follow the same approach as Rizzi and Ruggiero \cite{RizziRuggiero02,Ruggiero} and Rodrigues and  Sharif \cite{Rodrigues}, but reaching to a very different conclusion, because they use the Fermat metric (\ref{15}) as the metric of the space of reference instead of the euclidean metric (\ref{e16}).

In this case, the Fermat tensor is a projectable metric, therefore, it is really mathematically consistent to use  this metric as the metric of the space of reference, but although the disc does not suffer any deformation, when the rim of the disc is measured in the space of reference,  one obtains  $L' = \frac{2\pi R}{\sqrt{1-\omega^2R^2}}$ and $R'=R$. The physical interpretation of that fact \cite{Ruggiero} is that the rods along the rim contract, while the circumference does not; neither the rods along the radius nor the radius itself contracts.

In our opinion,  if neither the circumference itself nor the radius contracts, all the problems are generated by the rods used, precisely because this metric does not satisfies the free mobility axiom. Moreover,  it does not make sense to state that rods along the rim and along the radius have the same length if the free mobility condition is not assumed,  as far as this assertion is based on the free mobility axiom.

We claim that instead of $\hat g$, the space metric must be $\bar g$ (\ref{16}),the projectable metric that satisfies the free mobility axiom. As a  consequence,  the space of the disc is  Euclidean, and then $2\pi R^{\prime }=L^{\prime }=L=2\pi R$, so the lengths do not contract.  In this sense, our interpretation is closer to that of Klauber (see for instance \cite{Klauber98, Klauber99, Klauber00}).

The election of the Fermat tensor as the space metric is physically based on the Einstein convention of simultaneity, and the associated measuring  protocol, that  is  based on the assumption of the isotropy and constancy of the speed of light (See section \ref{sec:3}). Of course, this is true when local measurements are taken into account but this is not a so well-established fact in  global measurements. Think for instance about the Sagnac effect, that states that the transit time for a light ray to go around a closed path enclosing a non-null area depends on the sense of the curve followed by the light ray. 

Leaving aside some interpretations \cite{Tartaglia, RizziTartaglia98}\footnote{In these papers they claim the isotropy of the speed of light, locally and globally measured. To reach this conclusion an ad hoc correction in time measurements must be introduced, however, in \cite{Klauber03}  a good criticism to this approach can be found.},  in our opinion, this behaviour shows  that in global measurements the speed of light is anisotropic. This anisotropy is predicted even when the Fermat tensor is used to compute space distance,  as it can be seen in the above mentioned references \cite{RizziRuggiero02, Ruggiero, Rodrigues}, this apparent contradiction arising from the impossibility of synchronizing clocks around the rim. In fact,  measuring the global length of the rim using $\hat g$ is different from calculating the  sum of infinitesimal local measurements using locally the same tensor\footnote{The most interesting result of these approaches is that they  show that there are no contradictions between Special Relativity and General Relativity.  We  absolutely agree with them, and consequently, we think that it is not really justified the attempt to find an alternative to SRT exposed in some works of Selleri and Klauber, see for instance \cite{Selleri, Klauber98}.}.

\section{Anisotropy of the speed of light \label{sec:3}}

In previous sections we have claimed that the metric $\bar g$ (\ref{e8}) is the space metric of a given rigid reference frame $({\cal E}_3,\overline{g},\theta)$,  which must be used to measure space distances because it describes a specific physical reality, that is, the reality of the reference bodies used as standard rods. 


Even though we have eliminated the interpretation of the Fermat tensor $\hat g$  as the ``physical spatial metric'', this object has not lost its importance. On the contrary, it can be interpreted as the ``optical metric''. 

Consider two close points $P_1$ and $P_2$, and  consider also the three events 
$$e_1 =(x^i+dx^i, x^0+dx^0_{(-)}),\qquad e_2 = (x^i,x^0),\qquad e_3 =(x^i,x^0+dx^0_{(+)})$$
where, $e_1$ is the event on $P_1$ when a light signal is sent from $P_1$ to $P_2$, $e_2$ is the event when the signal arrives at $P_2$ and is instantaneously reflected to $P_1$ where it arrives at event $e_3$.

Since for  light signal $ds^2=0$, solving from
$$ds^2 = g_{\mu\nu}dx^\mu dx^\nu = g_{00}dt^2+2g_{0i}dx^0dx^i+g_{ij}dx^idx^j$$ 
we obtain the two solutions
\beq\label{e20}
dx^0_{(\pm)} = \frac1{g_{00}}\left(-g_{0i}dx^i\pm\sqrt{(g_{0i}g_{0j}-g_{00}g_{ij})dx^idx^j}\right)
\eeq

It is well known \cite{Landau} that physical time (that one measured by a clock in $P_1$)  during a closed trip is given by  
\beq \label{e21}
d\tau = \sqrt{-g_{00}}\delta x^0
\eeq
where $\delta x^0$ is lapse of the time coordinate between the departure and the arrival.
It is clear that, in the trip described,  $\delta x^0 = dx^0_{(+)}-dx^0_{(-)}$, so 
\beq\label{7.1}
d\tau =\frac2{ \sqrt{-g_{00}}}\sqrt{(g_{0i}g_{0j}-g_{00}g_{ij})dx^idx^j}=2 \sqrt{\hat{g}_{ij}dx^idx^j}\nonumber
\eeq
where we have introduced the Fermat metric in adapted coordinates.
\begin{remark}
The Fermat metric must be used to compute times spent by light in  infinitesimal round  trips.
\end{remark}

In traditional approaches, this fact and the assumption of the isotropy of the speed of light, implies that Fermat metric $\hat g$ can be used as the space metric, but as it has been  discussed before, we have considered  the flat metric $\bar g$ obtained by constriction as the space metric. Other possible relationships between $\hat g$ and $\bar g$ have also been considered in the literature: Bel's Meta-rigid motion, see for instance  \cite{BelLlosa, Bel99, Bel96},  and Fermat holonomic \cite{Ferhol} congruences, among others. 

Although on theoretical grounds preferring one to another is a matter of taste, the choice of $\overline{g}$ has measurable consequences. Indeed, given a rigid reference frame, there are two distinguished spatial metrics: first, the one belonging to the reference frame $\overline{g}$ , which is materialized by the reference body and is rigid, and the second one, the Fermat metric $\hat{g}$, which is embodied in radar signals  and is not generally preserved.

Thus, experiments can be devised to compare these two metrics: those used for measuring the anisotropy of the speed of light in vacuum. In other words, it is possible to check whether $\hat{g}$ and $\overline{g}$ are constrictional to each other.  

According to our interpretation, the distance between $P_1$ and $P_2$ is
\beq\label{7.2}
dl = \sqrt{\bar{g}_{ij}dx^idx^j}\nonumber
\eeq
therefore, the {\em one-way speed} of the signal is  
\beq\label{7.3}
v=\sqrt{\frac{\bar{g}_{ij}dx^i dx^j}{\hat{g}_{ij}dx^i dx^j}} \nonumber
\eeq

Let be $\{\bar{n}_{i(a)}\}$, with $ a= 1,2,3$, an orthonormal triad of $\bar{g}_{ij}$, that obviously defines a basis of the space, that is
\beq\label{7.4}
\bar{g}_{ij}= \delta^{ab}\bar{n}_{i(a)}\bar{n}_{i(b)}
\eeq

If we consider
 
\beq\label{7.5}
dx^i = \gamma^a\bar{n}^i_{(a)} dl \quad {\rm with}\quad \bar{n}^i_{(a)} = \bar{g}^{ij}\bar{n}_{i(a)}\quad\mbox{ and}\quad \delta_{ab}\gamma^a\gamma^b = 1
\eeq
then from (\ref{7.3}) we obtain
\beq\label{7.6}
v=\frac{1}{\sqrt{\hat{g}_{ij}\gamma^a\bar{n}^i_{(a)}\gamma^b\bar{n}^j_{(b)}}}
\eeq

Taking into account (\ref{e8}) in adapted coordinates, i.e.
\beq\label{7.7}
\bar{g}_{ij} = \exp(\theta)(\hat{g}_{ij}+\epsilon \mu_i \mu_j)
\eeq
where $\exp(\theta) = \varphi$, and writing $\mu$ in the above mentioned basis $\{\bar{n}_{i(a)}\}$,
\beq\label{7.8}
\mu_i = M^a\bar{n}_{i(a)}
\eeq
it is easy to see that the Fermat metric may be written as 
\beq\label{7.9}
\hat{g}_{ij}= \hat{c}^{ab}\bar{n}_{i(a)}\bar{n}_{i(b)}\quad{\rm with}\quad 
\hat{c}^{ab} = \exp(-\theta)\delta^{ab}-\epsilon M^a M^b
\eeq
and then (\ref{7.6}) yields 
\beq\label{7.11}
v=\frac{1}{\sqrt{\hat{c}^{ab}\gamma_a\gamma_b}} \quad {\rm where}\quad \gamma_a = \gamma^a
\eeq

so 
\beq\label{7.11is}
v^{-1} = \sqrt{\exp(-\theta)-\epsilon(\vec{M}\cdot\vec{V})^2}
\eeq
where $\vec{V}$ points the signal direction of motion. Hence, we conclude that 

\begin{remark}\label{r2}
In any point of the space, the speed of light is isotropic in the plane orthogonal to $\mu$.
\end{remark}

\subsection{Sagnac effect}

Taking into account (\ref{8.6}) and (\ref{e20}), if we consider a trip given by the parametric equations $r = R =constant, z = constant$, we obtain 
\beq
\frac{d\theta}{dx^0} = -\omega \pm\frac1R
\eeq
Integrating it, it is possible to find the coordinate time elapsed in $P_1$ from departure until the signal arrival, and we obtain 
$$x^0_{\pm} = \frac{2\pi R}{1\mp\omega R}$$
where the sign depends on whether the light travels, in the same sense (+) as the rotation or in the reverse (--). It is clear that all measures are taken in the same point $P_1$ of the platform. Using (\ref{e21}), we calculate  the physical time elapsed between the two arrivals, and we obtain
\beq\label{e30}
\Delta\tau = \sqrt{1-\omega^2R^2}(x^0_{+}-x^0_{-}) = \frac{4\pi R^2 \omega}{\sqrt{1-\omega^2R^2}}
\eeq
As it can be sought in \cite{Klauber02} or in \cite{RizziRuggiero03}, this expression yields,  in a first order approximation, the well known expression
$$\Delta Z = \frac{4\omega\cdot S}{\la c}$$
where $\omega$ is the constant angular velocity, $S$ is the vector associated to the area enclosed by the light path and $\la$ is the wavelength of light, as seen by an observer at rest on the rotating platform.

There are many interpretation of physical consequences of  (\ref{e30}):
 for instance, in \cite{RizziRuggiero03} they achieved to explain this effect imposing the uniformity of the speed of light, but they require the two ways path to have different lengths. However,  in \cite{Rodrigues} an anisotropy of the speed of light is predicted. Finally, in \cite{Klauber02} Klauber claims that this effect proves the failure of the Special Theory of Relativity.

Our interpretation is different:  the physical times spent by light in doing their round trips are
\beq \label{e31}
\tau_{\pm} = \sqrt{1-\omega^2R^2}x^0_{\pm}
\eeq
while the length of the periphery, as measured with the metric (\ref{e16}), is $l_{\pm} =\pm 2\pi R$, so we expect an anisotropy of the speed of light

\beq
c_{\pm} = \frac{-\omega R\pm 1}{\sqrt{1-\omega^2R^2}}=\pm \frac{\sqrt{1\mp\omega^2R^2}}{\sqrt{1\pm\omega^2R^2}}.
\eeq

The  times (\ref{e31}) coincide with the ones obtained by Klauber, and Rodrigues and Sharif in the above cited works, but for the last authors, the periphery of the disc, as commented in section \ref{sec:2.1} must be computed with the Fermat tensor (\ref{e17}), and then $l_{\pm} =\pm \frac{2\pi R}{\sqrt{1-\omega^2R^2}}$ so they predict an anisotropy 
$$c_{\pm} = \frac{1}{\omega R\pm 1}.$$ 

In conclusion,  an accuracy of  second order in ($\omega R$) is needed to decide between both approaches.

\subsection{The speed of light in a co-moving frame with the Earth \label{sec:7.3}}

In this section, following the way of Bel et al. in \cite{Bel99,  BelMol00, Aguirregabiria, BelMarMol94},   we will consider an experiment to test the local anisotropy of the speed of light using optical interferometers, as those made by  Michelson and Morley \cite{Michelson}. This will be done in a neighbourhood of the Earth's surface, and considering only the exterior field.

We consider, in the framework of General Relativity, the linear approximation of the gravitational field exterior to the Earth taking into account its mass ($M$), its mean radius, ($R$) and its reduced quadrupole moment ($J_2$).

As we will see, in the frame of reference co-moving with the Earth the local anisotropy of the space is of the order of $10^{-13}$, slightly smaller  than Bel's predictions ($2.5\times10^{-12}$ ), and closer to experimental issues ($2.1\times 10^{-13}$).

The line-element is given by \cite{BelMol00}:

\begin{equation}
\label {7.18}
ds^2=-(1-2U)dt^2+2A_idx^idt + (1+2U_G)\delta_{ij}dx^idx^j
\end{equation}
where $U = U_G +U_\Omega $ with
 \bea
\label {7.19}
U_G&=&\frac{M}{r}\left(1+\frac{1}{2}\frac{J_2R^2(1-3\cos^2 \theta
)}{r^2}\right)\\
U_\Omega &=& \frac12\Omega^2 r^2 \sin^2\theta
\eea
and where the angular velocity $\Omega$ is introduced via 
\beq
A_i = \Omega_{ij}x^j \quad {\rm with}\quad \Omega_{ij}=\epsilon_{ijk}\delta^k_3\Omega
\eeq

Of course (\ref{7.18}) is only an approximation, and only  terms proportional to  $U_G$ and $\Omega^2$ are considered.


Consider in this space-time the killing congruence given by 
\beq\label{7.20}
u = \frac{1}{\sqrt{1-2U}}\partial_t\approx(1+U)\partial_t
\eeq
then,  the Fermat tensor is
\beq\label{7.21}
d\hat s^2= \hat g_{ij}dx^idx^j = ((1+2U_G)\delta_{ij} + A_iA_j)dx^idx^j
\eeq
and it is easy to see that taking $\epsilon = -1$, and introducing the 1-form $\mu$
\bea\label{7.23b}
\mu &=& A_i(1+U)dx^i
\eea
that
\beq
g_{ij}+u_{i }u_{j}-\mu_i \mu_j  = (1+2U_G)\delta_{ij}dx^idx^j
\eeq
so this is equivalent to choose in (\ref{7.7}) 
\beq
\exp(-\theta) = 1+2U_G \Rightarrow \theta =-2U_G \label{7.23c}
\eeq

As in spherical coordinates:
\beq
x = r \sin(\theta)\cos(\phi), \;\;y = r \sin(\theta)\sin(\phi), \;\; z= r \cos(\theta)
\eeq
(\ref{7.23b}) becomes
\beq\label{7.24}
\mu = A_\phi(1+U)d\phi\quad {\rm on}\quad A_\phi = \Omega r^2 \sin^2(\theta)
\eeq 
therefore, we have the following remark: 

\begin{remark}
The space direction $\mu$, ``privileged'' in the Earth co-moving frame is at any point tangent to the corresponding  parallel and pointing to the east.
\end{remark}

Finally, we can introduce (\ref{7.23b}) and (\ref{7.23c}) into (\ref{7.11}) and compute the expected anisotropy. It follows, at the required approximation, that  (\ref{7.11}) yields to 
\beq\label{7.12}
v=1+\frac12\theta - \frac12 M^aM^b\gamma_a\gamma_b
\eeq

Since the interferometer  is usually kept horizontal, 
$$\gamma^1 = \cos A, \quad \gamma^2 = \sin A  \quad {\rm and}\quad \gamma^3 = 0$$
where $A$ is the azimuth of the direction $\gamma^a$, so we obtain
\bea\label{7.29}
v &=&1+ \frac12\theta -  \frac14 \left[M_1^2+M_2^{2} + (M_1^2-M_2^{2})\cos2A  + 2M_1M_2\sin2A\right] \nonumber\\
&=& a_0+ a_2\cos2A + b_2\sin2A
\eea 
So the  parameter characterizing the anisotropy is $\sqrt{a_2^2+b_2^2}$. In \cite{Brillet} a value of $2.1\times 10^{-13}$ was obtained for this quantity. In our case we have $\sqrt{a_2^2+b_2^2} =\frac14(M_1^2+M_2^2)=  \frac14\Omega^2R^2\sin^2(\theta)$ and taking into account the following values: $R = 6378164\,m$,  $\Omega = 2.434\times10^{-13}\,m^{-1}$  
at a colatitude $50^\circ$ yields to $3.6\times 10^{-13}$.

As mentioned above, this calculation is very similar to that shown in \cite{BelMol00}, where a value  of $2.5\times10^{-12}$ was obtained, and a value of $6.2\times10^{-13}$ when a global space-time is considered, that is, when  not only the external gravitational field but also the internal field is considered.

\section{Conclusions and outlook}

In this paper we recall theoretical results exposed in \cite{LlosaSoler04} and we squeeze their physical consequences. In particular, this allows us to give a new approach to the  Ehrenfest paradox, showing that no contradiction arises, because as   the rotating disc is not an inertial frame, it is  not possible to use Special Theory of Relativity results. As mentioned above there is a great number of references that explain mathematically the rotating rigid disc, but we  claim  that these explanations, instead of clarifying the physical behaviour of approximately rigid solids, suggest that Fermat tensor $\hat g$ is not the suitable metric to describe the rods used in the non time-orthogonal frames, that  any suitable metric must be projectable, but also, must have a constant curvature.

The theorem \ref{t3} states that in any spacetime, it  is always possible to find  rigid motions, and we interpret the constant curvature projectable metric $\bar g$, obtained by constriction from $\hat g$ as the appropriate one to  modelize the real rods used in a laboratory, while the Fermat metric must be used as an optical tensor to compute times spent by light in infinitesimal round trips.

This physical interpretation of $\bar g$ and $ \hat g$ allows us to examine the problem of the isotropy of the speed of light, and provide an explanation of the experimental result obtained by Billet and Hall, and also of the Sagnac effect. Moreover, our result is slightly different from the analogous ones present in the literature so, it would be desirable to perform  new experiments to decide which is the best metric for the space of reference.

In all the particular cases of theorem \ref{t3}, we have easily obtained the metric of the space because of  the symmetry of the problems, but in a future work the problem of modelizing an spinning top with arbitrary rotation will be considered, using perturbation theory in the partial differential system appearing in \cite{LlosaSoler04}.

\ack
The author is grateful to 
Josep Llosa and Bartolomé Coll, for valuable suggestions and comments.

%
%

\end{document}